\documentclass[aps,prl,superscriptaddress]{revtex4}

\usepackage{epsfig}
\usepackage{bm}
\usepackage{amsmath,amssymb, amsthm}
\usepackage{graphicx}
\usepackage{epsf}
\newcommand{\be}{\begin{equation}}
\newcommand{\ee}{\end{equation}}
\newcommand{\ba}{\begin{eqnarray}}
\newcommand{\ea}{\end{eqnarray}}
\newcommand{\nn}{\nonumber \\}

\begin{document}
\narrowtext

\title{Decoherence induced deformation of the ground state in adiabatic quantum computation}

\author{Qiang Deng}
\author{Dmitri V.~Averin\footnote{Corresponding author; e-mail: dmitri.averin@stonybrook.edu}}
\affiliation{Department of Physics and Astronomy, Stony Brook
University, SUNY, Stony Brook, NY 11794-3800 }

\author{Mohammad H.~Amin}
\author{Peter Smith}
\affiliation{D-Wave Systems Inc., 100-4401 Still Creek Drive,
Burnaby, B.C., Canada V5C 6G9}
\affiliation{Department of Physics, Simon Fraser University, Burnaby, British Columbia, Canada V5A 1S6}

%
\begin{abstract}
Despite more than a decade of research on adiabatic quantum computation (AQC), its decoherence properties
are still poorly understood. Many theoretical works have suggested that AQC is more robust against
decoherence, but a quantitative relation between its performance and the qubits' coherence properties,
such as decoherence time, is still lacking. While the thermal excitations are known to be important sources
of errors, they are predominantly dependent on temperature but rather insensitive to the qubits' coherence.
Less understood is the role of virtual excitations, which can also reduce the ground state probability
even at zero temperature. Here, we introduce normalized ground state fidelity as a measure of the
decoherence-induced deformation of the ground state due to virtual transitions. We calculate the normalized
fidelity perturbatively at finite temperatures and discuss its relation to the qubits' relaxation and
dephasing times, as well as its projected scaling properties.

\end{abstract}

\maketitle

Adiabatic quantum computation (AQC) \cite{farhi,gs}, either in its universal form \cite{ahar04,miz06},
or in the form of adiabatic quantum optimization \cite{exp1,exp2}, or quantum simulations \cite{sim},
presents a viable alternative to gate-model quantum computation (GMQC). Although a part of the original
motivation for introduction of the AQC \cite{gs} was the promise of the increased stability against
decoherence due to the energy gap between the ground and excited states, the question of the role of
decoherence in AQC remains an open one. This uncertainty makes it important to quantify more precisely the
decoherence properties of AQC. A crucial step towards this would be to define a quantitative characteristic
of the decoherence strength in AQC, that plays a role similar to the decoherence time for GMQC. However,
in the case of AQC, decoherence has qualitatively different, static effect on the qubits, not limiting
the operation time of an algorithm \cite{Amin09b}.

In AQC, adiabatic evolution of the ground state of a qubit system realizes the solution of a computational
problem represented by an appropriately designed Hamiltonian, which is typically written as
 \be
 H_S = A(s) H_D + B(s) H_P, \label{HS}
 \ee
where $s=t/t_f$ with $t_f$ being the total evolution time. At $s=0$, one has $A(0)=1$, $B(0)=0$,
and the system is initialized in the ground state of the initial (driver) Hamiltonian $H_D$, which usually
consists of the uniform superposition of all computational basis states. The energy scales $A(s)$ and $B(s)$
are then varied monotonically so that at $s=1$, $A(1)=0$ and $B(1)=1$. If the evolution is slow enough, an
isolated qubit system stays in the ground state with high fidelity throughout the evolution, and at $s=1$
reaches the ground state of the final (problem) Hamiltonian $H_P$, which provides a solution to a
computational problem.

If the qubit system is  weakly coupled to a dissipative environment, two effects are expected. First, the
low-frequency part of the environmental noise moves the system energy levels relative to each other. This
results in a dephasing of the energy eigenstates that eventually suppresses all off-diagonal elements of
the qubit density matrix in the energy basis. However, since the population of the ground state is the
only important part of the computation and the relative phases of the energy eigenstates do not carry any
information, this does not affect AQC. The second effect of the coupling to environment is that it induces
thermal transitions between the qubit energy levels pushing the qubit system towards thermal equilibrium
at a temperature $T$. In the limit of large $t_f$, the instantaneous probability to be in the ground state
asymptotically approaches the Boltzmann distribution, and so the qubit system loses some of the ground state
probability due to thermal occupation of the excited states. Such a thermal loss of probability can be compensated
by multiple iterations of an AQC algorithm as long as it does not scale exponentially with the size of the system,
i.e., as long as the number of excited states within roughly the energy $k
_BT$ above the ground state does not
grow exponentially.

The preceding arguments provide an intuitive explanation for the predicted robustness of AQC against local
environmental noise in the limit of weak coupling \cite{Lloyd,Amin09b,Amin09,Amin08,Tiersch,Sarandy,Roland,Childs}.
When the strength of the coupling to the environment is increased without changing either the Hamiltonian
or the temperature, the qubit system's Boltzmann distribution is still not directly affected. However, it is known
that the decoherence time of the qubits decreases with increased coupling, and strong coupling to the
environment eventually makes the qubits completely incoherent, rendering them useless for quantum computation.
In GMQC, qubit decoherence leads to computation errors which, without error correction, completely destroy
the computation process. This is why the qubits' quality factor, which is the ratio of the decoherence time
and the gate operation time, provides a good measure of the qubit performance in GMQC. It is, however, unclear
how an increase in coupling to the environment, or equivalently decrease in qubit quality factor, affects AQC.

In this paper, we look closely at what happens to the eigenstates of the qubit system in AQC when coupling to
the environment is non-negligible but still small enough to allow perturbation expansion. We introduce the
normalized ground state fidelity, defined as the distance between the open and closed system reduced density
matrices normalized to the Boltzmann ground state probability,  as a quantitative measure of decoherence-induced
deformation of the ground state in AQC, analogous to the decoherence time for GMQC. We calculate the fidelity
perturbatively and express it through the same environmental noise correlators that determine the decoherence
times in GMQC. Such an equilibrium calculation of the normalized ground state fidelity is accurate in the long
$t_f$ limit, but becomes approximate when the evolution is too fast for the system to reach local equilibrium.
However, the deviation from the equilibrium distribution is largest when the rate of relaxation between the
eigenstates becomes extremely small (e.g., near the end of the evolution in adiabatic quantum optimization).
As we shall see, the normalized ground state fidelity is closely related to the relaxation rate and becomes
close to 1, independent of the detailed probability distribution, when the relaxation is very slow. Therefore,
our calculation of the normalized fidelity can provide a good approximation in all regions as long as the
evolution time is not too short.

\section{Results}

\subsection{Normalized ground state fidelity - definition}

We first provide a definition for the normalized ground state fidelity based on the notion of fidelity between
two density matrices. To ensure consistent notation throughout this paper, symbols with (without) ``$\sim$''
denote quantities related to the coupled (uncoupled) qubit system and environment. We use letters $m,n$ to
enumerate the eigenstates and eigenvalues of the qubit system (e.g., $|n\rangle$, $E_n$), letters $\nu,\mu$
to enumerate the eigenstates and eigenvalues of the environmental degrees of freedom, and letters $a,b$ to
enumerate the eigenstates and eigenvalues of the {\em total} system (qubits+environment). The total Hamiltonian is
$\widetilde H=H_S+H_B+H_I$, where $H_B$ and $H_I$ are the environment and interaction Hamiltonians, respectively.
In the absence of coupling, $H_I{=}\,0$, and the eigenstates of the total system are $|a\rangle = |n\rangle
{\otimes} |\nu\rangle$ with eigenvalues $E_a=E_n+E_\nu$. When $H_I\ne 0$, the new eigenstates are $|\widetilde
a\rangle$, which typically are entangled superpositions of the unperturbed states $|a\rangle$. For weak coupling,
$|\widetilde a\rangle$ is very close to $|a\rangle$ and the effect of the environment is thermalization of the
qubit system. Once the environment is averaged out, the equilibrium of the total system gives the Boltzmann
distribution for the qubits:
\be
P_n = \sum_\nu {e^{-(E_n+E_\nu)/T} \over Z_SZ_B} = {e^{-E_n/T} \over Z_S}, \label{Boltzmann}
\ee
where $Z_S = \sum_n e^{-E_n/T}$ and $Z_B = \sum_\nu e^{-E_\nu/T}$ are the partition functions of the qubit
system and the environment.

As the coupling strength increases, the deviation of $|\widetilde a\rangle$ from $|a\rangle$ grows. In
equilibrium, the density matrix of the total system still has the Boltzmann form $\widetilde \rho_{SB} =
\sum_a \widetilde P_a |\widetilde a\rangle \langle \widetilde a|$, where $\widetilde P_a =
e^{-\widetilde E_a/T}/\widetilde Z_{SB}$, with $\widetilde Z_{SB} = \sum_a e^{-\widetilde E_a/T}$ being
the partition function of the total system. However, the reduced density matrix $\widetilde \rho_S =
\text{Tr}_B [\widetilde\rho_{SB}]$ of the qubit system alone is no longer given by the Boltzmann distribution.
The deviation from the Boltzmann form provides a good qualitative measure of how strongly the eigenstates
$|\widetilde a\rangle$ are deformed in comparison to the unperturbed states.

To quantify the loss of fidelity of the ground state due to such deformation of the energy eigenstates of the
uncoupled system, it is convenient first to separate this effect from the loss of fidelity due to thermal
excitations. This can be done by using {\em normalized} ground state fidelity, which we define as the Uhlmann
fidelity \cite{Uhlmann} between the reduced density matrix $\widetilde \rho
_S$ and the ``ideal'' ground state
density matrix $\rho_0=|0\rangle\langle 0|$, normalized to the Boltzmann ground state probability $P_0$:
\be
F = P_0^{-1/2} \text{Tr}\sqrt{\sqrt{\rho_0}\,\widetilde \rho_S\sqrt{\rho_0}} = \sqrt{\widetilde P_0/P_0}\, ,
\label{FDef} \ee
where $\widetilde P_0 = \langle 0|\widetilde \rho_S|0 \rangle$ is the equilibrium probability for the qubits
to be in the ideal ground state when coupled to the environment. Normalization to the {\em equilibrium}
Boltzmann ground state probability $P_0$ is natural in the context of this work, since in the calculations
presented below we adopt the assumption that the qubit-environment system maintains equilibrium throughout
the AQC evolution. In the weak-coupling limit, no deformation of the eigenstates is expected. Then
$\widetilde P_0 = P_0$, and Eq.~(\ref{FDef}) gives $F=1$. This shows that Eq.~(\ref{FDef}) indeed correctly
separates the effect of the quantum deformation of the ground state, which can be viewed as the result of
virtual transitions to the excited states, from the thermal loss of probability.
Qualitatively, the effect of the virtual transitions, expressed in $F$, is different from that of the
thermal transitions in two important aspects. First, it persists even at $T
=0$, when all the thermal
transitions are suppressed. Second, it depends on the strength of coupling to the environment (or decoherence
time of the qubits), while thermal equilibrium probabilities only depend on the energy eigenvalues
and temperature. Nevertheless, similarly to the thermal transitions, the virtual transitions reduce the
occupation probability of the ground state by transferring it to other low energy states. In general,
beyond the AQC, deviation of the occupation probabilities of an equilibrium quantum system from the Boltzmann
distribution due to non-vanishing strength of coupling to environment has been studied before as introducing
corrections to classical thermodynamics (see, e.g., \cite{AN,Nag02,Scully03,Nori10,WLJ} and references therein).
In particular, deformation of the ground state by coupling to an environment, described in our case as a
suppression of fidelity, is known to lead to several physical effects, e.g., suppression of the persistent
current in normal-metal rings \cite{PC} or violation of the fluctuation-dissipation theorem for the thermal
conductance \cite{FDT}.

\subsection{Perturbative calculation of fidelity}

We calculate the fidelity (\ref{FDef}) perturbatively and relate it to measurable parameters
of the qubit system and environment. As appropriate for AQC, we assume that the coupling $H_I$ is weak. This
allows us to employ the perturbation theory in $H_I$ around the non-interacting state of the qubit system
and the environment. To separate the effects of environment from other deviations from the perfect adiabatic
evolution of an AQC algorithm, we consider the limit of slow evolution, when the rate of change of the
control parameter of the Hamiltonian is small, e.g., smaller than the energy relaxation rate of
the qubit system. In this case, non-adiabatic transitions out of the ground state can be neglected, and the
occupation probabilities of the excited states (if non-vanishing) correspond to local equilibrium at each
moment during the evolution. While this is obviously not the most general case of time evolution in AQC, which
can be dominated by non-equilibrium effects, it covers the most important regime of ideal AQC evolution, and is
appropriate for the situation of sufficiently slow evolution in the presence of small but finite interaction
strength with environment we want to describe. The zeroth-order state of the perturbation theory in this regime
is the state with both the uncoupled qubit system and environment in equilibrium at the same temperature $T$,
i.e., the total density matrix of the system being the product of the density matrices $\rho_S=
\sum_n P_n |n\rangle\langle n|$ and $\rho_B = \sum_\nu P_\nu |\nu\rangle\langle\nu|$, where $P_n$ and $P_\nu$
are the Boltzmann probabilities.

In general, the reduction of the ground state probability due to finite $H_I$ is caused by two effects.
First is the change $\delta P_0$ in the equilibrium probability $P_0$ ($\equiv P_{n=0}$) as a result of
renormalization of the energy eigenvalues (Lamb shifts) of the qubit system. Second effect, conceptually more
important for this work, is the probability transfer into and out of the ground state due to renormalization
of the qubit system wavefunctions. Explicitly, the probability $\widetilde P_0$ that defines the normalized
fidelity (\ref{FDef}) can be expressed as
\be
\widetilde P_0 \equiv \widetilde P_{n=0} = \text{Tr}_{B,S} \Big[ |0\rangle\langle 0| \sum_a \widetilde P_a |
\widetilde a\rangle \langle\widetilde a|\Big] \, .
\label{p0} \ee
Introducing interaction-induced corrections to the equilibrium probabilities $\widetilde P_a=P_a+\delta P_a$,
where $P_a = P_n P_\nu$, and wavefunctions: $|\widetilde a(n,\nu) \rangle = |n\rangle {\otimes} |\nu\rangle+
|\delta \widetilde a(n,\nu) \rangle$, we can rewrite this expression to the lowest non-vanishing order in
$H_I$ as
\be
\widetilde P_0 = \delta P_0 + \sum_n P_n \text{Tr}_{B,S} \big[ |0\rangle\langle 0|{\otimes} \rho_B \cdot
|\widetilde a (n,\nu)\rangle \langle\widetilde a (n,\nu)| \big] \, .
\label{p0p} \ee
Using the relation $|0\rangle\langle 0| = 1 - \sum_{m\ne 0}|m\rangle\langle m|$ to transform the $n=0$ term
in Eq.~(\ref{p0p}) we obtain
\be
\widetilde P_0 = P_0 + \delta P_0 -  \sum_{n\ne 0} \left( \Gamma
_{0n} P_0 - \Gamma_{n0} P_n \right) \, ,
\label{FDB}
\ee
where
\[ \Gamma_{mn}\equiv \langle n|\text{Tr}_B [|\delta \widetilde a (m,\nu)\rangle \langle \delta
\widetilde a (m,\nu)|\rho_B ] |n\rangle \, .\]
The terms proportional to $\Gamma$ in Eq.~(\ref{FDB}) describe the reduction of the ground state probability
as a result of virtual transitions between the ground and excited state due to the interaction with the environment.

Next, we calculate $\delta P_0$ and $\Gamma_{mn}$. Quite generally, the interaction Hamiltonian $H_I$ is
\be
H_I = \sum_{j,\alpha} q^\alpha_j \sigma^\alpha_j,
\label{coupl}  \ee
where $\sigma^\alpha_j$ are the Pauli matrices for the $j$th qubit, $\alpha=x,y,z$, and $q^\alpha_j$ are
the corresponding operators of the noise generated by the environment. As usual, the averages of the noise
operators vanish, $\langle q^\alpha_j \rangle=0$. Then, in the weak coupling regime, the effect of
environment is fully characterized by the noise spectral densities:
\be
S^\alpha_j(\omega)=\int dt \ e^{i\omega t} \langle q^\alpha_j(t) q^\alpha_j(0) \rangle\, , \label{sp}
\ee
where $\langle ... \rangle= \text{Tr}_B\{\rho_B...\}$ is the average over the environmental degrees of freedom.
For simplicity, we limit our discussion to the most typical case when the noises with different $\alpha$ or
$j$ are uncorrelated. It is shown in the supplementary information (SI) that the perturbation expansion in $H_I$
in this situation gives
\ba
\delta P_0 &=& -\beta P_0 \sum_{j,\alpha,n,m} (P_n{-}\delta_{n0})|\sigma^\alpha_{j,nm}|^2 \int {d\omega \over 2\pi}
{S^\alpha_j(\omega) \over \omega_{mn} + \omega}\, , \nn
\Gamma_{mn} &=& \sum_{j,\alpha} |\sigma^\alpha_{j,nm}|^2 \int {d\omega\over 2\pi} {S^\alpha_j(\omega) \over
(\omega_{nm}+\omega)^2}\, ,
\label{Gammamn} \ea
where $\sigma^\alpha_{j,nm}\equiv \langle n|\sigma^\alpha_j|m\rangle$ and $\omega_{nm} \equiv E_n {-} E_m$.
Substituting (\ref{Gammamn}) into (\ref{FDB}) and then into (\ref{FDef}), we obtain
\ba
F =1 - \beta \sum_{j,\alpha,n,m} |\sigma^\alpha_{j,nm}|^2 \int {d\omega \over 4\pi} { S^\alpha_j(\omega)
(P_n{-}\delta_{n0}) \over \omega_{mn} + \omega} \nn - \sum_{j,\alpha,n>0}  |\sigma^\alpha_{j,n0}|^2 \int
{d\omega\over 4\pi} {S^\alpha_j(\omega)- (P_n/P_0)S^\alpha_j(-\omega) \over (\omega_{n0}+\omega)^2}.
\label{CentralEq} \ea
Equation (\ref{CentralEq}) is our main result. The normalized fidelity is well-defined at $T=0$, when all thermal
excitations are suppressed, i.e., $P_n=0$ for $n>0$ and $S^{\alpha}_j (\omega) \equiv 0$ at $\omega<0$.
Hence, the values of $\omega$ around $-\omega_{m0}$, when the denominator in (\ref{CentralEq}) vanishes,
do not contribute to the integral. When $T\neq 0$, the divergences that appear at $\omega= -\omega_{m0}$
reflect the fact that environment can also create real thermal excitations of the qubit system. However,
the detailed balance relation, $S^\alpha_j(-\omega)= e^{-\beta \omega}S^\alpha_j(\omega)$, ensures that
these divergences cancel each other out and Eq.~(\ref{CentralEq}) is well-defined also at $T\neq 0$ (see the SI).

\subsection{Normalized fidelity for single qubit}

Equation (\ref{CentralEq}) is now applied to specific problems. The first example we consider is a typical
{\em individual qubit} with the Hamiltonian
\be
 H_S=-[\epsilon \sigma^z+\Delta \sigma^x]/2
\label{sq} \ee
coupled as in Eq.~(\ref{coupl}), but only through $\sigma^z$, to the environmental noise with spectral density
$S(\omega)$ (\ref{sp}). In the usual weak-coupling approximation [see, e.g., Ref.~\onlinecite{blum}], the qubit
decoherence time $T_2^*$ is given by
\be
{1 \over T_2^*} = {1 \over 2T_1} + {1 \over T_\varphi}\, ,
\ee
where $T_1$ and $T_\varphi$ are the relaxation and pure dephasing times, given by
\ba
T_1^{-1} &=& (\Delta^2/\Omega^2)[S(\Omega) + S(-\Omega)]\, , \label{T1} \\
T_\varphi^{-1} &=& (\epsilon^2/\Omega^2)S(0)\, , \label{T2}
\ea
with $\Omega = \sqrt{\Delta^2+\epsilon^2}$. The standard expressions for the eigenstates of the Hamiltonian
(\ref{sq}) reduce Eq.~(\ref{CentralEq}) for the normalized fidelity to $F
= 1 - \Delta^2 K/2\Omega^2$, where
\be K=\int \frac{d\omega S(\omega{-}\Omega)}{2\pi \omega} \left\{ \frac{ 1{-}e^{- \omega/T}}{\omega} -
\frac{e^{-\Omega/T}{+}e^{- \omega/T}}{ T(e^{-\Omega/T}{+}1)} \right\} .
\label{K} \ee
We see that the same noise spectral density that defines the relaxation and dephasing rates (\ref{T1})
and (\ref{T2}) of the qubits in the GMQC determines the reduction of the ground-state normalized fidelity in AQC.
In this respect, the main difference between the reduction of normalized fidelity and the real-time relaxation and
dephasing is that even in the lowest-order perturbation theory, the normalized fidelity is reduced by the whole
spectrum of environmental excitations, and not just by limited spectral groups resonant with the
qubit energy differences or the low-frequency excitations, as in Eqs.~(\ref{T1}) and (\ref{T2}).

\begin{figure}[t]
\includegraphics[width=8.5cm]{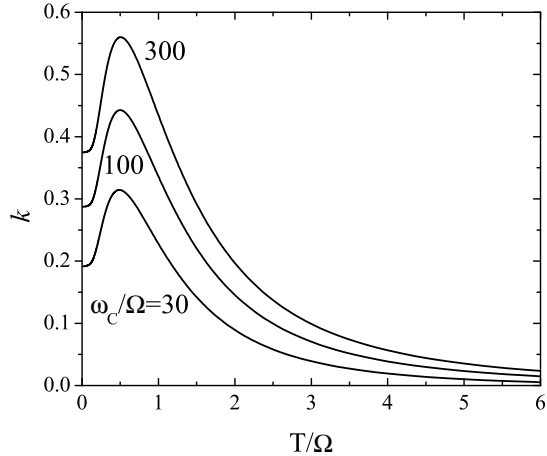}
\caption{The temperature-dependent factor $k$ in the expression \protect (\ref{OhmicF}) for the normalized ground
state fidelity of an individual qubit in the presence of Ohmic environment with cut-off frequency
$\omega_c$. }
\label{fig1} \end{figure}

To strengthen this comparison, we consider an Ohmic bath characterized by the noise
\be
S(\omega)=\eta\omega/(1-e^{-\omega/T}) \label{SOhmic}
\ee
where, $\eta$ is a dimensionless coefficient and $\omega_c$ is the cutoff frequency. In this case, the relaxation time is $T_1^{-1} = \eta(\Delta^2/\Omega)
\coth (\Omega/2T)$ and the normalized fidelity is expressed as
\be
F = 1 - \frac{k}{Q} \, , \;\;\;\;\; k\equiv {K \over 2\eta}  \tanh \frac{\Omega}{2T} \, ,
\label{OhmicF} \ee
where $Q=T_1\Omega$ is the qubit quality factor due to relaxation. Equation (\ref{K}) gives the
following expressions for the factor $k$ at low and high temperatures:
\be
k ={1\over 4\pi} \left\{ \!   \begin{array}{ll}  \ln (\omega_c/ \Omega) - 1 +
\pi^2T^2 /3\Omega^2\, ,  & T\ll \Omega ,\\  (\Omega/ T)^2 \ln (\omega_c /T), & T\gg \Omega \, .
\end{array} \!\!\! \right.
\label{hf}  \ee
Equation (\ref{OhmicF}) relates the normalized ground state fidelity to the qubit quality factor, $Q$, as calculated
due to relaxation only. This shows that the normalized fidelity can be related more closely to the relaxation ($T_1$)
and not dephasing ($T_\varphi$) processes. Adding a $1/f$ low-frequency noise of a realistic magnitude
does not change this conclusion (as discussed in more details in the numerical examples below). As
expected, a larger $Q$ leads to a better $F$. Figure~\ref{fig1} shows the factor $k$ in
Eq.~(\ref{OhmicF}) as a function of temperature for different cut-off frequencies $\omega_c$. It exhibits
the non-monotonic $T$-dependence, and only weak, logarithmic, dependence on $\omega_c$, which allows one
to estimate $F$ without precisely specifying $\omega_c$. The factor $k$ is maximal around $k_{\rm max}
\simeq 0.5$ at $T\simeq 0.5 \Omega$, which leads to a minimum normalized fidelity $F \simeq 1- (2Q)^{-1}$. Notice
that even a qubit quality factor as low as $Q=10$, which is practically useless for GMQC, leads to $F>95$\%.


\subsection{Normalized fidelity for multi-qubit systems}

We now consider multi-qubit systems, starting with a system of $N$ {\em uncoupled} qubits. In this case,
the trace in the definition of normalized fidelity (\ref{FDef}) can be taken independently over separate qubits, so
that the total $F$ is the product of fidelities $F_j$, $j=1, ..., N$ of the individual qubits:
$F=\prod_j F_j$. For instance, a typical starting point of AQC algorithms is to initialize the system in the
ground state of the Hamiltonian $H_D$ (\ref{H}). Then, the state of all qubits is the same and can be
characterized by the same normalized fidelity (\ref{OhmicF}). Then,
\be
F = (1-k/Q)^N \Big|_{Q\gg k} \simeq e^{-k N/Q} . \label{FUncoupled}
\ee
For independent qubits, $F$ scales exponentially with $N$ as a result of the exponential
scaling of the probability for all qubits to remain in their corresponding ground states. Since $Q$ is
inversely proportional to the noise strength $\eta$, by decreasing the noise by a factors of, e.g., 10,
one can achieve the same $F$ with 10 times more qubits.

\begin{figure}[t]
\includegraphics[width=9.5cm]{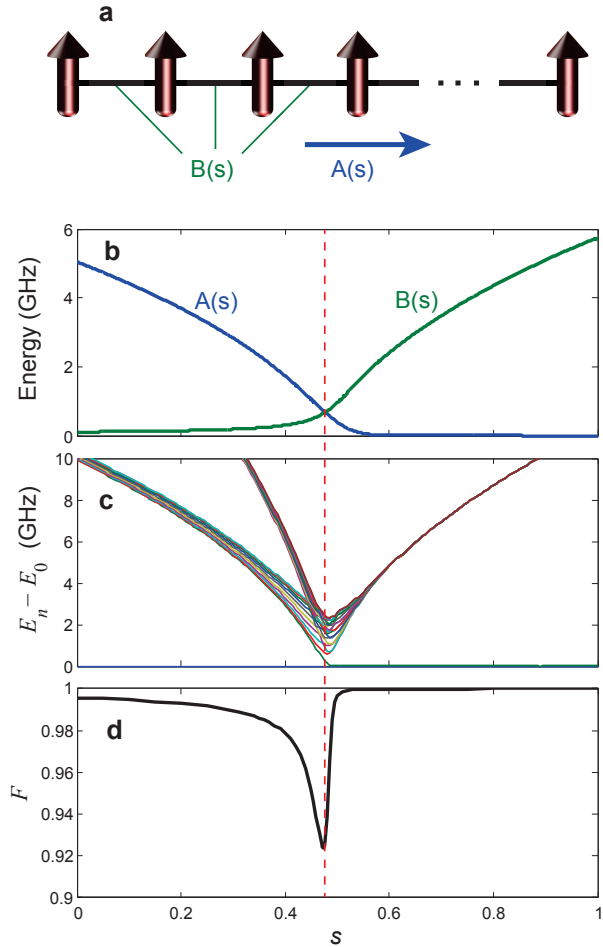}
\caption{{\bf a.} A ferromagnetic spin chain with transverse field and coupling energies given, respectively, by
$A(s)$ and $B(s)$ in Eq.~(\ref{HS}). {\bf b.} Energy scales $A(s)$ and $B(s)$ extracted from experimental parameters.
{\bf c.} The lowest 20 energy levels, relative to the ground state, of a 10-qubit ferromagnetic chain with $J_{ij}=-1$,
as a function of the normalized time $s$. {\bf d.} Normalized ground state fidelity of the 10-qubit chain of {\bf c} at
$T=20$ mK. The vertical (red) dashed line marks the quantum critical point as defined by the condition
$A(s_c){=}\,B(s_c)$.}
\label{fig2} \end{figure}

Next, we focus on how the normalized ground state fidelity behaves in practical AQC systems. We use as an example the
{\em D-Wave One} quantum annealing processor, as the one installed at the University of Southern California
(see Ref.~\onlinecite{exp2}). The Hamiltonian implemented by the processor has the form of Eq.~(\ref{HS}), with
\be
H_D = -\sum_{i=1}^N\sigma^x_i\, , \quad
H_P = \sum_{i=1}^Nh_i\sigma^z_i + \sum_{i,j=1}^{N}J_{ij} \sigma^z_i\sigma^z_{j}\, ,
\label{H} \ee
where $h_i$ and $J_{ij}$ are tunable dimensionless bias and coupling coefficients. The parameters $A(s)$ and
$B(s)$ for this processor are plotted in Fig.~\ref{fig2}{\bf b}. We calculate $F(s)$ for a ferromagnetic chain
(illustrated in Fig.~\ref{fig2}{\bf a}) with $h_i{=}\,0$ and $J_{i,i+1}{=}-1$, otherwise known as a quantum
Ising model in a transverse field. Here, the length of the chain is varied from $N{=}\,2$ to 16. Although this
model is exactly solvable (see, e.g., Ref.~\onlinecite{is} and references therein), $F$ cannot be calculated
exactly for practical noise models in which the coupling to environment is dominated by the $\sigma^z_j$ terms.
Hence, we calculate the normalized fidelity numerically. In the limit $N{\to}\infty$, the model is known to have
a {\em quantum critical point} at $s_c$ where $A(s_c){=}\,B(s_c)$. At this point, the chain goes through a quantum
phase transition between quantum paramagnetic and ferromagnetic phases. In the ferromagnetic phase, the ground state
is doubly degenerate with respect to simultaneous change of signs of all $\sigma^z_i$ terms. Figure \ref{fig2}{\bf c}
plots several of the lowest energy levels of a 10-qubit chain relative to the ground state energy $E_0$. In the
thermodynamic limit ($N\to \infty$), the appearance of the doubly-degenerate ground state and the minimum in the
energy gap between the ground and the second excited states happen at the quantum critical point. For the 10-qubit
chain of Fig.~\ref{fig2}, however, these happen at slightly different points than the one defined by
$A(s_c){=}\,B(s_c)$.

To calculate $F(s)$ for this system, we use a realistic noise model relevant to the {\em D-Wave} qubits \cite{Harris10}.
In this case, the dominant environmental coupling is to the magnetic flux noise, which couples directly to the qubit
computational basis states represented by the $\sigma^z_j$ operators. The noise spectral density $S(\omega)$ was
characterized in the earlier experiments, which were consistent with the noise being a combination of the $1/f$
low-frequency noise and an Ohmic noise at high frequencies \cite{env2}. For calculations of $F$, we take
$S(\omega)=\kappa(s)[S_{HF}(\omega) + S_{LF}(\omega)]$, where $S_{HF}(\omega)$ is the Ohmic spectral density
(\ref{SOhmic}) and $S_{LF}(\omega)=\gamma^2/|\omega|$. The coefficient $\kappa(s)=B(s)/B(s_m)$ appears because the
strength of coupling to flux noise depends on the persistent current of the flux qubits which changes as a function
of $s$ (see SI). Here, $s_m$ is the bias point at which the measurements of $\eta$ and $\gamma$ are performed. Based
on the experimental data, we use $\eta=0.1$, $\gamma = 20$ MHz and $s_m = 0.636$. We also assume $\omega_c=100$
GHz for the high-frequency cutoff and $\omega_L = 1$ MHz for the low-frequency cutoff (based on a $t_f{\sim} 1~\mu$s
evolution time of an algorithm). We found that for these parameters, $F$ is dominantly determined by the
high-frequency Ohmic noise and not by the $1/f$ noise.

In principle, since the total number of energy levels grows exponentially with $N$, the time required for numerical
calculation of $F$ also grows exponentially. Fortunately, the value of $F$ converges rapidly for a finite number
of retained energy states. Here, we keep all energy levels for $N\leq 10$, and up to 2000 energy levels for larger
chains. The normalized ground state fidelity $F(s)$ of the 10-qubit chain is plotted as a function of $s$ in
Fig.~\ref{fig2}{\bf d}. The fidelities of chains with other lengths (and also coupled systems other than chains)
are qualitatively the same as the one plotted in Fig.~\ref{fig2}{\bf d}. It is clear from the figure that $F(s)$
is minimum close to the critical point $s_c$. Notice also that the fidelity approaches 1 as $s\to 1$, which is
the result of $H_P$ commuting with $H_I$, with only $\sigma^z_j$ terms and negligible other types of coupling
to environment. This again reflects the fact that $F$ depends rather on relaxation than dephasing.

\begin{figure}[t]
\includegraphics[width=9.5cm]{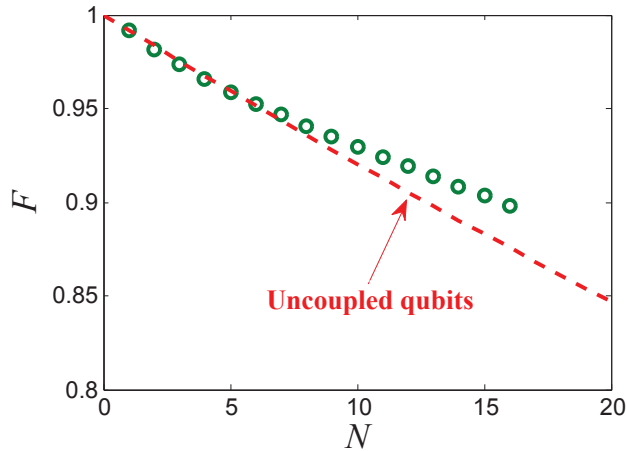}
\caption{Normalized ground state fidelity at the quantum critical point for ferromagnetic chains with $N=1$ to 16, at $T=20$ mK.
Circles are the numerical results using (\ref{CentralEq}). The red dashed line is the fidelity of uncoupled qubits from
(\ref{FUncoupled}), with $k=0.32$ and $Q=38.4$.} \label{fig3}
\end{figure}

Figure \ref{fig3} shows the numerical results for the normalized ground state fidelity for $N$-qubit chains with $N=1$
to 16 at the critical point. For all chain lengths, $F(s_c)$ is better than 90\%. It should be emphasized that these
are the minimum fidelities at the quantum critical point $s_c$. The normalized fidelity at all other points is larger,
and near $s=1$, is very close to 1 as shown in Fig.~\ref{fig2}{\bf d}. We have also plotted in Fig.~\ref{fig3} the
normalized ground state fidelity of $N$ uncoupled qubits (with the same parameters) at different $N$ based on the
exponential scaling of Eq.~(\ref{FUncoupled}). The scaling and magnitude of $F$ at large $N$ is better for the
ferromagnetic chain than for the uncoupled qubits. A plausible reason for this is that the spin-spin interaction
introduces additional rigidity into the chain dynamics making it less susceptible to the environmental perturbations,
and therefore increasing the fidelity. Unfortunately, it was not possible to pursue numerical calculations
beyond 16 qubits, as direct perturbation approximation would break down when $F$ strongly deviates from unity. A naive
exponential extrapolation of the data points to $N=128$ (representing the worse case) still yields $F=0.47$, meaning
that the eigenstates could retain their quantum properties without error correction for such a large-size system. As
in uncoupled qubits, if one can reduce the noise by a large factor, the size of the chain can be increased by the same
factor while keeping $F$ unchanged. In addition, other techniques such as dynamical decoupling \cite{DDecoupling} or
error correction \cite{ECorrect} could be employed to enhance the normalized ground state fidelity at large scales.

\section{Discussion}

Finally, we discuss how the normalized ground state fidelity should affect the performance of AQC. First, we notice that
the actual equilibrium ground state probability at point $s$ is $\widetilde P_0(s) = P_0(s)F^2(s)$, where $P_0(s)$ is the
Boltzmann probability. Therefore, a suppression of the normalized fidelity creates an extra reduction of the ground state
probability on top of the thermal reduction. In universal AQC \cite{ahar04,miz06}, $F(1)$ directly affects the quality of
the computation. Indeed, deviations of $F(1)$ from 1 mean that the statistics of measurements done on the final state will
be different from the one that corresponds to the ideal ground state. For instance, in the case of one qubit with
$\epsilon=0$ and the Hamiltonian (\ref{sq}), measurement of $\sigma_x$ has a non-vanishing probability $1-F^2$ of
producing the result $\sigma_x=-1$ different from the ground state $\sigma_x=1$ even at temperatures $T\ll \Delta$.
However, this effect is absent in the special case when the coupling to environment via $H_I$ commutes with the
final Hamiltonian $H_P$, leading to $F=1$ at the end of evolution, as in the adiabatic quantum optimization discussed
above and shown in Fig.~\ref{fig2}{\bf d}. In this case, a small fidelity in the middle of the evolution increases
the loss of probability due to the thermal transitions, thereby decreasing the ground state probability even
further. Therefore, the probability will be distributed among the low energy states
even more than implied by the thermal equilibrium. Part of the probability can be regained later when the gap is
larger and $F$ is closer to 1. However, since the relaxation time becomes exponentially long near the end of
evolution, the majority of the probability that is lost may not be gained back, thus leading to a smaller
probability of success. This makes it important to maintain $F(s)$ close to unity throughout the evolution.
We stress that most treatments of AQC based on the weak coupling master equation, e.g.,
\cite{Childs,Amin09b,master_eq}, even with Lamb shift, do not take into account the effect of deformation of the
eigenstates that is captured by our calculation of the normalized fidelity.

In summary, we have proposed using normalized ground state fidelity as a quantity for measuring the strength of
decoherence effects in AQC. The fidelity plays a role similar to decoherence time in GMQC, but takes into account
qualitatively different effects of environment on the ground state relevant to AQC. The fidelity is related to the
relaxation processes and is relatively insensitive to the dephasing. Our numerical calculations indicate that a
normalized fidelity close to unity can be achieved with a moderate qubit quality factor, even for large numbers of
qubits. Normalized ground state fidelity should be a useful measure of the environment related quality of AQC systems
in the context of further work on important topics in AQC such as quantum error correction or the threshold theorem.

\subsection{Acknowledgements}

MHA is grateful to A.J.~Berkley, M.W.~Johnson, R.~Liu, T.~Mahon, F.~Nori, A.~Yu.~Smirnov, and B.~Wilson for
discussions and comments.

\subsection{Author contributions}

All authors contributed equally to all aspects of this work.

\subsection{Competing financial interests}

The authors declare no competing financial interests in relation to this work.

\end{document}